\documentclass[traditabstract]{aa}

\usepackage{txfonts,natbib,graphicx}
\bibpunct{(}{)}{;}{a}{}{,}

\def\ltsima{$\; \buildrel < \over \sim \;$}
\def\simlt{\lower.5ex\hbox{\ltsima}}
\def\gtsima{$\; \buildrel > \over \sim \;$}
\def\simgt{\lower.5ex\hbox{\gtsima}}
\def\igr{J00291}
\def\igrall{\object{IGR J00291+5934}}

\def\saxj{\object{SAX J1808.4--3658}}
\def\rxte{\it RXTE}

\begin{document}

\title{Spin down during quiescence of the fastest known accretion-powered  pulsar}

\author{A. Papitto\inst{\ref{inst1}}$^{,}$\inst{\ref{inst2}} \and A.Riggio \inst{\ref{inst2}}$^{,}$\inst{\ref{inst1}} \and L.Burderi \inst{\ref{inst1}} \and T.Di Salvo \inst{\ref{inst3}}\and A.D'A\'i \inst{\ref{inst3}}\and R.Iaria \inst{\ref{inst3}}}

\institute{Dipartimento di Fisica, Universit\'a degli Studi di Cagliari, SP Monserrato-Sestu, KM 0.7, 09042 Monserrato, Italy\label{inst1}
\and
INAF - Osservatorio Astronomico di Cagliari, Poggio dei Pini, Strada 54, 09012 Capoterra (CA), Italy\label{inst2}
\and
Dipartimento di Scienze Fisiche ed Astronomiche, Universit\'a di Palermo,
 via Archirafi 36, 90123 Palermo, Italy\label{inst3}}

\abstract{ We present a timing solution for the 598.89 Hz accreting
  millisecond pulsar, {\igrall}, using Rossi X-ray Timing Explorer
  data taken during the two outbursts exhibited by the source on 2008
  August and September. We estimate the neutron star spin frequency
  and we refine the system orbital solution.  To achieve the highest
  possible accuracy in the measurement of the spin frequency variation
  experienced by the source in-between the 2008 August outburst and
  the last outburst exhibited in 2004, we re-analysed the latter
  considering the whole data set available. We find that the source
  spins down during quiescence at an average rate of
  $\dot{\nu}_{sd}=(-4.1\pm1.2)\times10^{-15}$ Hz s$^{-1}$.  We discuss
  possible scenarios that can account for the long-term neutron star
  spin-down in terms of either magneto-dipole emission, emission of
  gravitational waves, and a propeller effect.  If interpreted in
  terms of magneto-dipole emission, the measured spin down translates
  into an upper limit to the neutron star magnetic field, $B\simlt
  3\times 10^{8}$ G, while an upper limit to the average neutron star
  mass quadrupole moment of $Q\simlt2\times10^{36}$ g cm$^2$ is set if
  the spin down is interpreted in terms of the emission of
  gravitational waves.

}


\titlerunning{Spin down of the fastest known accreting pulsar}
\authorrunning{Papitto et al.}
\maketitle
\section{Introduction}

The discovery of the first accreting millisecond pulsar (AMSP) in
1998, {\saxj} \citep{WijvdK98}, confirmed the predictions of the
recycling scenario, according to which millisecond radio pulsars are
the end product of a long phase of accretion of matter and angular
momentum onto a neutron star (NS) hosted in a low mass X-ray binary
\citep[see, e.g.,][]{BhtvdH91}.  In the twelve years since the first
discovery, the class of AMSPs has grown to thirteen members, all X-ray
transients.  To perform a timing analysis of different outbursts of
the same source allows the estimate of its evolution over a time range
of a few years. In the case of {\saxj}, the observations of five
outbursts over 10 yr has allowed a firm estimate of its spin and
orbital evolution. The orbital period has been observed to increase at
a rate of nearly two orders of magnitude larger than what is predicted
by conservative mass transfer (\citealt{DS08,B09}, see also
\citealt{H08}, H08 hereafter).  This has led the authors to argue that
a large fraction of the mass transferred by the companion star is
ejected by the system taking away the angular momentum needed to match
the observed value.  A regular NS spin down has also been measured by
H08 \citep[see also][]{H09} leading to stringent upper limits on the
various mechanisms that can brake down a pulsar during quiescence such
as magneto-dipole emission, emission of gravitational waves and a
propeller effect. These effects, and in particular the spin down
torque associated with the emission of gravitational waves, $N_{GW}$,
crucially depend on the spin frequency of the NS ($N_{GW}\propto
\nu^5$). It is therefore very appealing to shed light on the long-term
behaviour of the fastest AMSP discovered so far, the 598.89 Hz pulsar
{\igrall} ({\igr} in the following). In this paper, we present a
timing analysis based on the two outbursts shown by the source in
2008, and observed by the Rossi X-ray Timing Explorer ({\it RXTE}).
The results thus obtained are compared with the rotational state of
{\igr} at the end of the outburst exhibited on 2004 December, that is
the only other outburst of this source for which high temporal
resolution data are available.

\section{Observations}
\label{sec:obs}
The X-ray transient, {\igr}, was discovered by INTEGRAL on 2004
December 2 \citep{S05}. The 598.89 Hz pulsations found in its light
curve make it the fastest AMSP discovered so far \citep[][G05
  hereinafter]{G05}.

\begin{figure}
\resizebox{\hsize}{!}{\includegraphics{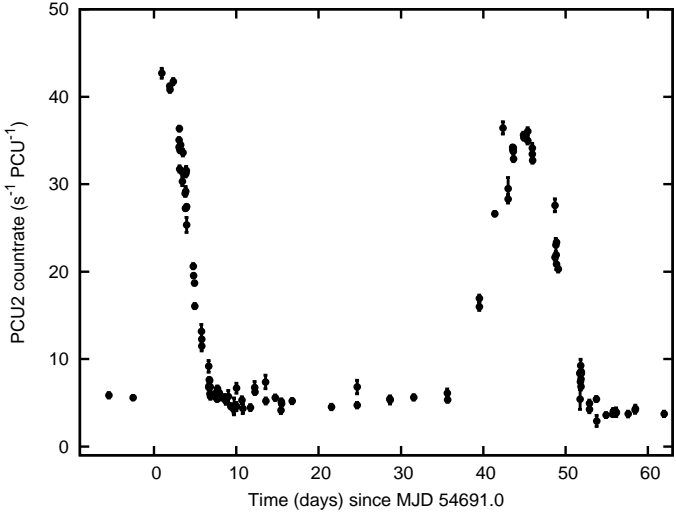}}
\caption{Lightcurve of the two outbursts exhibited by {\igr} during
  2008, as observed by the PCU2 of the PCA aboard {\rxte}.  }
\label{fig:lc}
\end{figure}

Renewed activity was detected by {\it RXTE} on 2008 August 13
\citep{C08}.  The 2.5--25 keV X-ray flux\footnote{The spectrum of
  {\igr}, as observed by the PCA aboard {\rxte}, is evaluated by
  modelling data recorded by the top layer of the PCU2 with an
  absorbed power law. We fix the nH to $0.43\times10^{22}$ cm$^{-2}$
  \citep{Pzs05}. A 6.4 keV iron line is sometimes needed to model the
  spectrum.}  reaches a peak level of $(6.3\pm0.2) \times 10^{-10}$
erg cm$^{-2}$ s$^{-1}$, which is $\approx 0.5$ times the peak flux
observed during the 2004 outburst (G05). The flux decreases on a
timescale $\tau\approx3$ d and the source returns to quiescence
$\sim5$ d after the first detection. The light curve recorded by the
PCU2 of the Proportional Counter Array (PCA) aboard {\rxte} is plotted
in Fig.\ref{fig:lc}. As the nearby source V709 Cas (17 arcmin away)
contributes to the X-ray flux detected by {\rxte} in the direction of
{\igr} \citep{M08}, the observed count-rate stays at a level of $\sim
6$ c s$^{-1}$ PCU$^{-1}$ (corresponding to $(7\pm2)\times 10^{-11}$
erg cm$^{-2}$ s$^{-1}$; 2.5--25 keV) even when the {\igr} outburst is
presumably over.  {\igr} is again detected in outburst on 2008
September, 21 and the fluence of this second episode is similar to
that of the first one.

To perform a timing analysis on the 598.89 Hz pulsar signal, we
consider events recorded by the PCA (Obsid P93013) both in good xenon
(1$\mu$s temporal resolution), and event mode (125$\mu$s temporal
resolution) configurations.  All the arrival times were first
corrected with respect to the Solar System barycentre, considering the
position of the optical counterpart determined by \citet[][T08
  hereinafter]{T08}, RA=00$^h$ 29$^m$ 03$^s$.05 $\pm$ 0$^s$.01,
DEC=59$^{\circ}$ 34' 18''.93 $\pm$ 0''.05.  A re-analysis of the data
taken by {\rxte} during the 2004 outburst (ObsId P90052 and P90425),
is also reported. Despite a temporal analysis of the 2004 outburst of
{\igr} having already been performed by G05, \citet[][F05 in the
  following]{Fal05} and \citet[][B07]{B07}, such a re-analysis is
aimed at deriving the most accurate estimate of the spin frequency at
the end of the outburst that can then be compared to the spin
frequency of the source measured in 2008, after $\approx 3.7$ yr of
quiescence.

\section{Temporal analysis}

\subsection{The 2008 outbursts}
\label{sec:timing}

To check the presence of pulsations during the 2008 outbursts of
{\igr}, we first correct the photon-arrival times for the source
orbital motion. As no eccentricity was detected as a result of the
timing analysis performed on the data of the 2004 outburst (G05, see
also Sect. \ref{sec:2004}), we consider a circular orbit to correct
photon-arrival times, $t_{em}-t_{arr}=x \sin{[l(t_{em})]}$. Here
$t_{arr}$ and $t_{em}$ are the photon arrival and emission times,
respectively, $x=a\sin{i}/c$ the projected semi-major axis of the NS
orbit, { $l(t_{em})=2\pi(t_{em}-T^{*})/P_{orb}$ is the mean orbital
  longitude}, $P_{orb}$ the orbital period, and $T^{*}$ the epoch at
which the mean orbital longitude is equal to zero \citep[see][for a
  discussion of this choice of orbital epoch\footnote{The epoch of
    passage at the ascending node, $T_{asc}$, a fiducial in true
    longitude that has been widely used in the analysis of AMSPs, is
    related to T$^{*}$ by the relation,
    $T_{asc}=T^{*}+(P_{orb}/\pi)e\sin\omega$, where $e$ is the
    eccentricity and $\omega$ is the longitude of periastron measured
    from the line of nodes \citep{DeeBoyPrv81}.}]{DeeBoyPrv81}.  After
a set of values for the orbital parameters is considered, the
corrected emission times are obtained by iterating the above relation
until the difference between successive steps is of the order of the
RXTE absolute timing accuracy (3.4$\mu$s, \citealt{Jah06}).  We first
consider the values of $x$ and $P_{orb}$ given by G05 as orbital
evolution is not expected to change them significantly during the time
elapsed between the 2004 and 2008 outbursts ($\approx 3.7$ yr). The
propagation of the error in the value of T* quoted by G05 yields
instead an uncertainty of $\sim 80$ s.  We then use the technique
described by \citet{Pap05} to improve the estimate of T*.
\begin{figure}
\resizebox{\hsize}{!}{\includegraphics{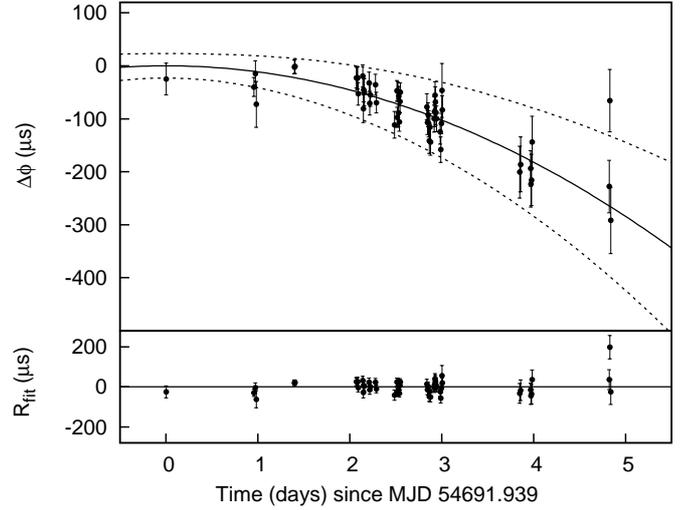}}
\caption{Evolution of the pulse phase delays (in $\mu$s) during the
  August 2008 outburst obtained by folding around the best estimate of
  the spin frequency at the beginning of that outburst,
  $\nu_{F}^{A08}=598.89213046$ Hz, the time series preliminarily
  corrected with the orbital solution listed in the left column of
  Table \ref{tab}. The solid line is the best-fit quadratic model,
  while dashed lines mark the 1$\sigma$ confidence-level
  intervals. The bottom panel shows residuals with respect to a model
  that also includes the spin up term.}
\label{fig:residuiA08}
\end{figure}
The difference of $\simeq -200$ s between the improved estimate we
find and the value predicted according to the G05 solution indicates
how a correction of $\simeq -0.015$ s (2.5$\sigma$ from the G05
estimate) to the value of the orbital period has to be applied.  Using
this improved orbital solution and folding 500s-long data segments in
12 phase bins around the frequency $\nu_F^{A08}=598.89213046$ Hz, we
detect pulsations at the 99\% confidence level in the interval MJD
54691.9--54696.8, which we refer to as the 2008 August
outburst. Pulses are again detected in the interval MJD
54730.5--54740.1 (2008 September outburst), after observations have
been folded around the frequency $\nu_F^{S08}=598.89213060$ Hz.  A
detection is assessed according to the criterion stated by
\citet{Leh83}, rejecting the profiles that have a probability larger
than 1$\%$ of being due to chance.

The pulse profiles have an rms amplitude of $\simeq 8\%$ and are
successfully modelled by a sinusoid.  We fit the phases thus evaluated
with the relation
\begin{equation}
  \label{eq:phases}
\phi(t)=\phi(0)+(\nu_0-\nu_{f})\;(t-T_{ref})+\frac{1}{2}\dot{\nu}\;(t-T_{ref})^2+R_{orb}(t),
\end{equation}
where $T_{ref}$ is the reference epoch for the timing solution,
$\nu_0$ and $\dot{\nu}$ are the pulsar frequency at the reference
epoch and its mean derivative across the outburst, respectively, and
$R_{orb}(t)$ describes the phase residuals due to a difference between
the parameters used to correct photon arrival times and the actual
orbital parameters of the system. Neglecting second-order terms in the
eccentricity, these residuals behave as 
{\setlength\arraycolsep{2pt}
\begin{eqnarray}
R_{orb}(t) & = & x \nu_f \left\{ \sin{[l(t)]} \;\frac{\delta x}{x} -\frac{1}{P_{orb}}[l(t)\;\delta P_{orb}+2\pi\;\delta T^*]\cos{[l(t)]}   \right\}+ {} 
\nonumber  \\
& &   {} +x \nu_f \left\{\frac{1}{2}sin{[2l(t)]} h - \frac{1}{2}cos{[2l(t)]}g \right \},
\end{eqnarray}
 where  $h=e\cos{\omega}$ and $g=e\sin{\omega}$, $e$ is
  the eccentricity of the orbit, $\omega$ the longitude of the
  periastron measured from the ascending node, and the terms $\delta
  x$, $\delta P_{orb}$, and $\delta T*$ are the differential
  corrections to the respective orbital parameter with respect to
  those used to correct the time series. If significant corrections to
  the orbital parameters are found, photon arrival times are corrected
  with the new set of orbital parameters and the phases thus obtained
  are again fitted using Eq. (\ref{eq:phases}). This procedure is
  iterated until no orbital residuals are significantly detected.

\begin{figure}
\resizebox{\hsize}{!}{\includegraphics{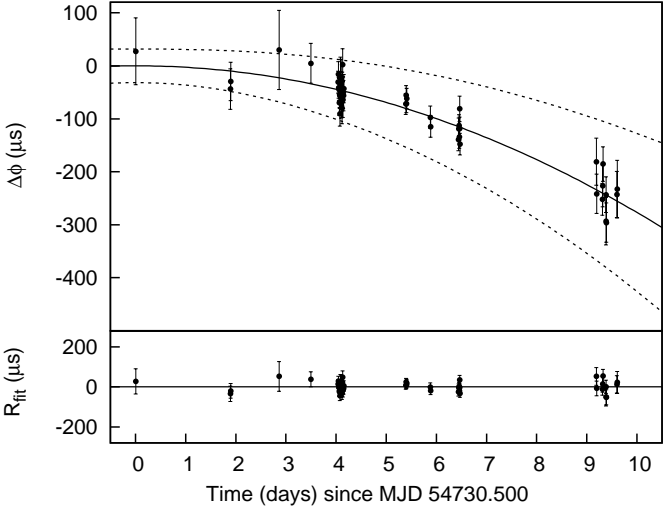}}
\caption{Same as Fig. \ref{fig:residuiA08} concerning the phases of
  the 2008 September outburst. Observations were folded around the best
  estimate of the spin frequency at the beginning of that outburst,
  $\nu_F^{S08}=598.89213060$ Hz.}
\label{fig:residuiS08}
\end{figure}

 The results we obtain by fitting the phases of the two outbursts
 separately are given in the leftmost and central column of Table
 \ref{tab}. We consider either a constant frequency model (i.e.,
 $\dot{\nu}=0$) and also allow for the possibility of a constant spin
 frequency derivative during each of the outbursts ($\dot{\nu}\neq0$
 model). The addition of a quadratic component to the fit of 2008
 August and 2008 September data does not significantly improve the
 model, and only an upper limit could be set on the spin up term
 during both outbursts. However, the results obtained with this model
 are considered as more reliable than those obtained putting
 $\dot{\nu}=0$, as a non-zero spin frequency derivative is expected on
 physical grounds and also already observed during the outburst shown
 by the source during 2004 (F05 and B07; see also
 Sect. \ref{sec:2004}). We note however that the spin frequency
 determined in the case $\dot{\nu}=0$ is compatible within a 3$\sigma$
 confidence level with that obtained allowing $\dot{\nu}\neq0$, thus
 representing a particular case of this more general solution.  The
 evolution of the phases determined for the 2008 August and September
 outbursts are plotted in Figs. \ref{fig:residuiA08} and
 \ref{fig:residuiS08}, respectively, together with the residuals with
 respect to the best-fit ($\dot{\nu}\neq0$) models.

 \begin{table*}
\begin{minipage}[t]{170mm}
\caption{Spin and orbital parameters of {\igr}.\label{tab}} \centering
\begin{tabular}{lrrr}
 \hline
& 2008 August & 2008 September  & 2004 December \\
\hline

$\Delta t$ \tablefootmark{a} & 54691.9 -- 54696.8 & 54730.5 -- 54740.1 & 53342.3 -- 53352.0  \\
T$_{ref}$ (MJD)\tablefootmark{b} &54691.939&54730.500& 53352.0 \\
\hline
a$\sin{i}$/c (lt-ms)&64.988(6) & 64.982(6) &64.990(1) \\
P$_{orb}$ (s) &8844.07(2) &8844.078(9)&8844.079(1) \\
T* (MJD) &54691.938749(5) &54730.529222(5)&53345.1619264(5) \\
e &$<7\times10^{-4}$ &$<6\times10^{-4}$&$<1.4\times10^{-4}$ \\
\hline
$\dot{\nu}=0$ model & & &\\
$\nu$ (Hz) &598.89213082(4) &598.89213082(2) & \\
$\chi^2$/dof &80.7/48&42.2/41 & \\
\hline
$\dot{\nu}\neq0$  model & & \\
$\nu$ (Hz)&598.89213046(13) &598.89213060(8) &  598.89213094(1)\\
$\dot{\nu}$  ($\times 10^{-13}$ Hz s$^{-1}$) &$<21$& $<4.5$ & $+5.1\pm0.3$\\
$\chi^2$/dof &68.9/47 &34.7/40 & 497.3/429  \\
\hline
\end{tabular}
\tablefoot{ Numbers in parentheses are 1$\sigma$ errors in the last
  significant digit. Upper limits are evaluated at 3$\sigma$
  confidence level.  The uncertainties have been scaled by a factor
  $\sqrt{\chi^2_r}$ to take into account a reduced $\chi^2$ of the
  best-fit model larger than 1. The uncertainties quoted in the
  estimates of $\nu$ and $\dot{\nu}$ do not include the systematic
  errors due to the positional uncertainty. \tablefoottext{a}{Time
    interval covered by the presented solution}
  \tablefoottext{b}{Reference epoch for the timing solution.}  }
\end{minipage}
\end{table*}

\subsection{The 2004 outburst}
\label{sec:2004}

 This work is mainly focused on the study of the spin down experienced
 by {\igr} during the quiescent phase that lasted since the end of the
 2004 outburst to the onset of the 2008 August outburst.  To this aim,
 the most accurate estimate of the spin frequency at the end of the
 2004 outburst is needed. Previous works relied only on a fraction of
 the available data: G05 considered only the first three days of data,
 ObsId P90052, while the solutions of F05 and B07 are valid for the
 subsequent seven days, ObsId P90425, the only set of which data were
 publicly available at that time. We re-analysed the 2004 outburst
 including all the available {\rxte} data (ObsId P90052 and
 P90425). Time series were corrected using the position of the optical
 counterpart (T08), while the position of the proposed radio
 counterpart \citep{Rpn04} was considered in previous works.  The
 results we obtained, applying the same procedure outlined in
 Sec. \ref{sec:timing} are listed in the rightmost column of Table
 \ref{tab}. Only the parabolic model ($\dot{\nu}\neq0$) is presented
 because the quadratic term is highly significant ($\Delta\chi^2=303$
 over 429 degrees of freedom with respect to a constant frequency
 model, which has one degree of freedom more). The timing solution is
 referred to an epoch at the end of the outburst to estimate the spin
 frequency after the accretion-induced spin-up is over, which can
 therefore be compared to the frequency at the beginning of the 2008
 August outburst. The orbital parameters we obtain are compatible with
 those previously published by G05, F05, and B07, and are somewhat
 more precise as they rely on a longer baseline. The addition of three
 days of data to the data set considered by F05 and B07 indicates an
 average spin-up term [$\dot{\nu}_{04}=(+5.1\pm0.3)\times10^{-15}$ Hz
   s$^{-1}$], which is lower with respect to those there evaluated
 [$(8.5\pm1.1)\times10^{-15}$ Hz s$^{-1}$]. The phase evolution and
 residuals with respect to the best-fit model are plotted in
 Fig.\ref{fig:residuiD04}.

\begin{figure}
\resizebox{\hsize}{!}{\includegraphics{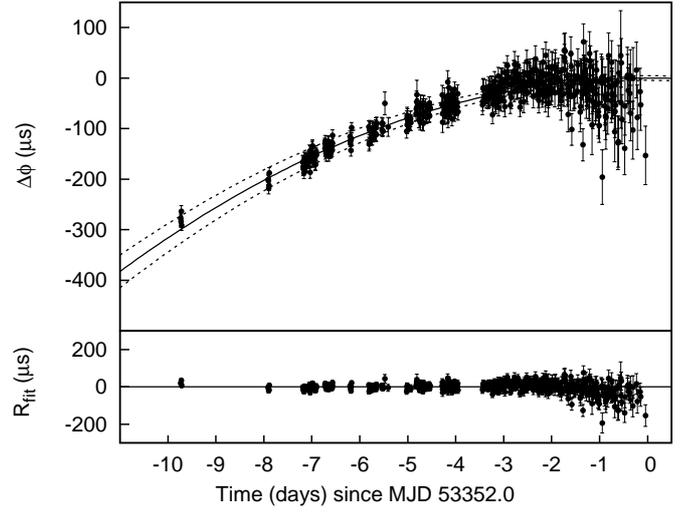}}
\caption{Same as Fig. \ref{fig:residuiA08} and
  Fig. \ref{fig:residuiS08} concerning the phases of the 2004
  outburst. The timing solution has been refereed to the epoch at the
  end of the outburst, MJD 53352.0, to get the most precise estimate
  of the spin frequency at that epoch, $\nu_F^{D04}=598.89213094$ Hz,
  around which the observations are folded.}
\label{fig:residuiD04}
\end{figure}

\subsection{The positional uncertainty}
\label{sec:pos}

The uncertainties quoted in Table 1 are 1$\sigma$ error based on the
modelling of the phase evolution. However, the error in the source position
introduces an uncertainty in the determination of the pulse phases
\begin{equation}
\Delta\phi_{pos}=\nu y [\sin(M_0+\epsilon)\cos\beta\delta{\lambda}-\cos(M_0+\epsilon)\sin\beta\delta{\beta}],
\end{equation}
where $y$ is the Earth distance from the Solar System barycentre in
lt-s, $\lambda$ and $\beta$ are the ecliptic longitude and latitude,
respectively, $\delta{\lambda}$ and $\delta{\beta}$ the respective
uncertainties, $M_0=[2\pi(T_0-T_{\gamma})/P_{\oplus}]-\lambda$, $T_0$
is the start time of observations considered, $T_{\gamma}$ is the
nearest epoch of passage at the vernal point, $P_{\oplus}$ is the
Earth orbital period, and $\epsilon=2\pi(t-T_0)/P_{\oplus}$
\citep[see, e.g., ][]{LynGsm90}. For time intervals that are small
with respect to the Earth orbital period (i.e. $\epsilon << 1$) such
as the ones considered in this work, this expression can be expanded
as a polynomial and only the lowest order terms retained. As the value
of the spin frequency at the reference epoch of the timing solution
depends on the linear term of the temporal evolution of the phases
[see Eq. (\ref{eq:phases})], the systematic uncertainty introduced by
the position error in this measure, $\delta\nu_{pos}$, is
\begin{equation}
\label{eq:errpos}
\delta\nu_{pos}\simeq \nu \;y \left(\frac{2\pi}{P_{\oplus}}\right)[\cos{M_0}\;\cos{\beta}\;\delta\lambda+\sin{M_0}\;\sin{\beta}\;\delta\beta].
\end{equation}
Considering the uncertainties in the position quoted by T08
[$\sigma_{\lambda}\leq(3.9\times10^{-5})^{\circ}$,
  $\sigma_{\beta}\leq(2.4\times10^{-5})^{\circ}$], and evaluating this
relation for the 2008 and the 2004 outbursts translates into
$\sigma_{\nu}^{08}\simeq 1\times10^{-8}$ Hz and
$\sigma_{\nu}^{04}\simeq 3 \times 10^{-8}$ Hz, respectively.  To get
reliable estimates of the uncertainties affecting each measured spin
frequency, these systematic errors have to be summed in quadrature
with the statistical errors quoted in Table \ref{tab}. The systematic
error in the frequency variation between the two outbursts can instead
be estimated as $\sigma_{\Delta\nu\;pos}\simeq4\times10^{-8}$ Hz.

\subsection{The spin evolution of {\igr} during quiescence}
\label{sec:evol}

The estimate of the spin frequency of {\igr} at the beginning of the
2008 August can be compared with the spin measured at the end of the
2004 outburst to measure the frequency variation experienced by the
source during quiescence. Considering the value measured using the
$\dot{\nu}\neq0$ model at the onset of the 2008 August outburst
($\nu^{A08}=598.89213046(13)$ Hz), we thus obtain $\Delta
\nu=\nu^{A08}-\nu^{04}=-(0.48\pm0.13\pm0.04)\:\mu$Hz, where the first
error quoted is the statistical error given by the difference of the
values quoted in Table \ref{tab} and the latter reflects the
uncertainty on the source position. The average spin-down rate during
quiescence is therefore estimated as
$\dot{\nu}_{sd}=(-4.1\pm1.1\pm0.3)\times10^{-15}$ Hz s$^{-1}$. The
large uncertainty affecting this estimate is due to the limited
statistics available for the 2008 August outburst which imply loose
estimates of the spin frequency at the beginning of that outburst and
of the spin frequency derivative during that outburst (see also the
discussion).

\section{Discussion and conclusions}
\label{sec:disc}

We have presented a detailed timing analysis concerning the two
outbursts shown by {\igr} in 2008 August and September as observed by
{\it RXTE}, as well as a re-analysis of the 2004 outburst using the
whole {\it RXTE} dataset available.

Our analysis of the 2004 data confirms the significant spin up the
source has underwent while accreting as already reported by F05 and
B07. The estimate of the spin up rate we have presented here
[$\dot{\nu}_{04}=(+5.1\pm0.3)\times10^{-13}$ Hz s$^{-1}$] is evaluated
on a longer temporal baseline with respect to that considered by those
authors and is accordingly more accurate. The magnitude of the spin up
is lower by a factor $\sim$0.4 with respect to that quoted by those
works, making its interpretation easier in terms of the NS accretion
of the supposedly Keplerian disc angular momentum at the flux emitted
by the source (see discussion in B07).

In contrast, no derivative is detected significantly during the 2008,
August and September, outbursts, with 3$\sigma$ upper limits on the
spin up component of $|\dot{\nu}_{A08}|<2\times10^{-12}$ and
$|\dot{\nu}_{S08}|<4.5\times10^{-13}$ Hz s$^{-1}$, respectively. These
estimates reflect the limited statistics available. An
accretion-induced spin-up is expected to depend almost linearly on the
mass accretion rate ($\dot{\nu}\propto \dot{M}^{1-\alpha/2}$, where
$\alpha$ is the index of the dependence of the inner disk radius on
the mass accretion rate, $R_{in}\propto\dot{M}^{-\alpha}$,
$\alpha=2/7$ if the inner disc radius is approximated by the Alfven
radius). As the peak X-ray flux shown by the source during both 2008
outbursts is roughly half that of the 2004 outburst, the spin up rate
is expected to scale accordingly, provided that the flux is a good
tracer of the mass accretion rate. We thus expect $\dot{\nu}\approx
2.5\times10^{-13}$ Hz s$^{-1}$ during each of the outbursts shown by
the source in 2008. While the upper limit to the September 2008
outburst spin up is of the same order as this value, the upper limit
to the August 2008 data is one order of magnitude larger and can
therefore not be considered as a tight constraint. This can also be
viewed by considering the 3 $\sigma$ upper limit to the difference
between the spin frequencies at the beginning of the 2008 September
and August outbursts, $\nu_{S08}-\nu_{A08}<0.45\:\mu$Hz.  Neglecting
any spin down in-between the two outbursts, the spin up during the
2008 August episode cannot be larger than $|\dot{\nu}_{A08}|\simlt
1\times10^{-12}$ Hz s$^{-1}$ to account for this difference. This
reasonable upper limit is already smaller by a factor of two than the
upper limit found from timing analysis of that outburst alone.

The comparison of the spin frequency measured at the beginning of the
2008 August outburst with that of the end of the 2004 episode
indicates that the spin frequency has decreased during quiescence.
Summing in quadrature the statistical error to the systematics induced
by the uncertainty in the source position (see Sec. \ref{sec:evol}),
we quote an average spin-down rate during quiescence of
$\dot{\nu}_{sd}=(-4.1\pm1.2)\times10^{-15}$ Hz s$^{-1}$.

A spin down at a rate of $(-5.5\pm1.2)\times 10^{-16}$ Hz s$^{-1}$
extending over $\sim 10$ yr, has already been measured by \citet[][see
  also H08]{H09}, from the 401 Hz AMSP, {\saxj}. Riggio et al. (2010,
in prep.) found an average spin down rate of
$(-5.5\pm1.2)\times10^{-15}$ Hz s$^{-1}$ for the case of \object{XTE
  J1751--305}, while only an upper limit ($|\dot{\nu}|\leq
2\times10^{-15}$ Hz s$^{-1}$, 95\% confidence level) could be set
instead by \citet{Pat10a}, for \object{SWIFT J1756.9--2508}.
Similarly to H08, we discuss the spin down measured from {\igr} in
terms of: (i) magneto-dipole radiation, (ii) emission of gravitational
waves, and (iii) the propeller effect.

The spin down luminosity of a rotating magnetosphere,
$L_{sd}=4\pi^2I\nu\dot{\nu}$, has been evaluated by \citet{Spt06} in
the limit of force-free magneto-hydrodynamics as,
$L_{sd}=(1+\sin^2{\alpha})\mu^2(2\pi\nu)^4/c^3$, where $\alpha$ is the
latitude of the magnetic poles, $I$ is the NS moment of inertia and
$\mu$ is the magnetic dipole. This translates into a spin down rate,
$\dot{\nu}_{sd}=L_{sd}/(4\pi^2 I \nu)=
[3(1+\sin^2{\alpha})/(2\sin^2{\alpha})] (N_{vac}/2\pi I)$, where
$N_{vac}= - (2/3) \mu^2 (2\pi\nu/c)^3 \sin^2{\alpha}$ is the usual
expression for the torque acting on a magnetised rotator in vacuum.
The estimate of $\dot{\nu}_{sd}$ we have given translates into a value
of the magnetic dipole of $\mu\simeq1.1(2)\times10^{26}
\;I_{45}^{1/2}\;(1+\sin^2{\alpha})^{-1/2}$ G cm$^{3}$, where $I_{45}$
is the moment of inertia in units of $10^{45}$ g cm$^2$. This
estimates translates into a magnetic field, $B\simlt
2.2(4)\times10^{8} \;I_{45}^{1/2}\;(1+\sin^2{\alpha})^{-1/2}$ G at the
magnetic poles of a 10 km NS. Considering $\alpha=0$, an upper limit
of $\simeq 3\times10^{8}$ G (3$\sigma$ confidence level) on the
magnetic field is obtained. This estimate fits well into the expected
range of magnetic field strengths for the AMSPs to be the progenitors
of recycled radio millisecond pulsars ($\simeq 10^8$--$10^9$ G). It is
also compatible with the requirements set on the dipole strength by
the maximum and minimum accretion rate experienced by the source while
showing pulsations.  For pulsations to be observed, the magnetospheric
radius has to lie between the NS radius, $R_{NS}$, and the corotation
radius, $R_{C}=(GM/4\pi^2\nu^2)^{1/3}$ ($23.6$ m$_{1.4}^{1/3}$ km for
{\igr}, where m$_{1.4}$ is the NS mass in units of 1.4
M$_{\odot}$). The minimum flux at which we observe pulsations during
the 2008 outbursts is $F_{2.5-25}=(1.8\pm0.4)\times10^{-10}$ erg
cm$^2$ s$^{-1}$ (MJD 54696.751). Assuming as the bolometric correction
factor that derived by G05 (2.54) and that the observed X-ray flux
reflects the mass accretion rate, this translates into
$\dot{M}_{min}\simeq0.9\times10^{-10}$ m$_{1.4}^{-1}$ R$_{10}$ d$_4^2$
M$_{\odot}$ yr$^{-1}$, where R$_{10}$ is the radius of the NS in units
of 10 km, and d$_4$ is the distance to the source in units of 4 kpc.
Considering the value quoted by G05 for the peak flux during the 2004
outburst, the maximum accretion rate at which pulsations were observed
can be estimated as $\dot{M}_{max}\simeq4.7\times10^{-10}$
m$_{1.4}^{-1}$ R$_{10}$ d$_4^2$ M$_{\odot}$ yr$^{-1}$.  Using the
expressions derived by \citet{PsaCha99}, the presence of pulsations at
these two limiting accretion rates indicates that the magnetic dipole
has to lie in the range, (0.2 -- 21)$\times 10^{26}$ d$_4$ G cm$^{3}$,
fully compatible with our estimate. As the minimum flux at which
pulsations are observed is likely overestimated by a factor of $\sim
2$ because of the contribution of V709 Cas, the upper limit to the
magnetic dipole is likely to be a factor $\sqrt{2}$ smaller.
Considering also the dynamical estimate of the maximum mass-accretion
rate derived by B07 from the spin up rate observed during the 2004
outburst, the lower limit to the dipole strength increases to
0.6$\times 10^{26}$ G cm$^{-3}$, still compatible with the estimate
derived here.  Our estimate of the magnetic field strength is also
compatible with the upper limit estimated by T08 from the X-ray
quiescent luminosity, $<3\times10^8$ G, using the criteria stated by
\citet{B02} and \citet{DSB03}.

The spin down torque associated with the emission of gravitational
radiation has been proposed to explain the non-detection of accreting
pulsars with frequencies higher than $\approx730$ Hz (\citealt{C03},
see \citealt{Wag84,Bld98,MelPay05} for models describing mechanism
that can lead to a non-zero mass quadrupole for an accreting pulsar).
In this case, the spin-down torque is, $N_{GW}=
-(32/5)GQ^2(2\pi\nu/c)^5$ \citep[see, e.g.,][]{Tho80}. Under the
hypotheses that the spin down of {\igr} is due only to this mechanism
and that the torque due to the GW emission is constant, our measure of
the average spin down translates into an estimate of the average mass
quadrupole moment, $Q\simeq1.2(2)\times10^{36}\;I_{45}^{1/2}$ g
cm$^2$. Considering the upper limit at the 3$\sigma$ confidence level,
$Q\simlt 2\times10^{36}$ g cm$^{2}$, the maximum amplitude at the
Earth of the emitted GW is therefore, $h_C\simlt
4.6G(2\pi\nu)^2Q/dc^4\simlt 3\times10^{-28}$ $d_4^{-1}$ $I_{45}^{1/2}$
\citep{Brd98}. Assuming that the spin down during quiescence of {\igr}
and {\saxj} is driven by the emission of GW and that the NS in these
systems have a similar mass quadrupole, the spin down driven by the
emission of GW should be $\approx (598.9/401.0)^5\simeq7.6$ times
larger in {\igr} than in {\saxj}. The large uncertainties affecting
the spin down estimates in both sources do not allow us to check
whether this prediction is compatible with observations. However, that
the spin down of both sources can be easily explained by
magneto-dipole emission of a NS with a magnetic field of the order of
that expected for an AMSP makes it unlikely that the spin down during
quiescence of AMSPs is dictated by the emission of GW.

The spin down of an accreting NS during quiescence can be also
explained by the propeller effect \citep{IllSun75}, that is the
centrifugal inhibition of accretion by a magnetosphere that extends
beyond the corotation radius.  Considering the upper limit at
$3\sigma$ on the average spin frequency derivative during quiescence
that we have measured ($\dot{\nu}_{sd}<-0.5\times10^{-15}$ Hz
s$^{-1}$), the system should eject matter at an average rate
$\dot{M}_{ej}\simgt 2\times10^{-12}$ $n^{-1}$ $(r_{in}/R_C)^{-1/2}$
I$_{45}$ m$_{1.4}^{-2/3}$ M$_{\odot}$ yr$^{-1}$, if the spin down is
explained in terms of the propeller effect alone. Here, $R_{in}$ is
the inner disc radius, and $n$ is the dimensionless torque
(\citealt{GL79}), which takes values $n\approx 1$, as soon as
$r_{in}\simgt R_C$ \citep{Eks05}. Assuming that mass is propelled away
from the NS at a roughly constant rate, the source quiescent
luminosity would then be $L_p\geq
GM\dot{M}_{ej}/2R_C\simeq6\times10^{33}$ erg s$^{-1}$ . As the
quiescent flux received from the source is $F_q\simlt 1.2\times
10^{-13}$ erg cm$^{-2}$ s$^{-1}$ \citep[0.5--10 keV,][]{Cam08,Jon08},
the source should be farther than $\simeq20$ kpc to match this value,
and a distance greatly in excess of 10 kpc is obviously to be excluded
(see also G05).  We thus conclude that it is highly unlikely that the
propeller effect alone explains the spin down of {\igr}.

Observations of future outbursts from this source will be used to
monitor the constancy of the long-term spin down, and to derive
tighter constraints on the parameters of the NS in {\igr}.

\begin{acknowledgements}
This work is supported by the Italian Space Agency, ASI-INAF
I/088/06/0 contract for High Energy Astrophysics, as well as by the
operating program of Regione Sardegna (European Social Fund
2007-2013), L.R.7/2007, ``Promotion of scientific research and
technological innovation in Sardinia''.

Soon after this paper was first submitted, other two papers discussing
the rotational evolution of this source during quiescence appeared on
arXiv.org \citep{Pat10b,Hrt11}. Even if the analysis presented by
these authors slightly differs with respect to that presented here [in
  particular \citet{Pat10b} derived a timing solution under the
  preliminary assumption that pulse phases linearly correlate with the
  X-ray flux], the values they obtain for the spin-down rate of the
source during quiescence are entirely compatible with that presented
here. We thank Jacob M.~Hartman for useful discussions and comments on
this paper.

\end{acknowledgements}

\bibliographystyle{aa}
\bibliography{14837}

\begin{thebibliography}{38}
\expandafter\ifx\csname natexlab\endcsname\relax\def\natexlab#1{#1}\fi

\bibitem[{{Bhattacharya} \& {van den Heuvel}(1991)}]{BhtvdH91}
{Bhattacharya}, D. \& {van den Heuvel}, E.~P.~J. 1991, \physrep, 203, 1

\bibitem[{{Bildsten}(1998)}]{Bld98}
{Bildsten}, L. 1998, \apjl, 501, L89+

\bibitem[{{Brady} {et~al.}(1998){Brady}, {Creighton}, {Cutler}, \&
  {Schutz}}]{Brd98}
{Brady}, P.~R., {Creighton}, T., {Cutler}, C., \& {Schutz}, B.~F. 1998, \prd,
  57, 2101

\bibitem[{{Burderi} {et~al.}(2007){Burderi}, {Di Salvo}, {Lavagetto}, {Menna},
  {Papitto}, {Riggio}, {Iaria}, {D'Antona}, {Robba}, \& {Stella}}]{B07}
{Burderi}, L., {Di Salvo}, T., {Lavagetto}, G., {et~al.} 2007, \apj, 657, 961

\bibitem[{{Burderi} {et~al.}(2002){Burderi}, {Di Salvo}, {Stella}, {Fiore},
  {Robba}, {van der Klis}, {Iaria}, {Mendez}, {Menna}, {Campana}, {Gennaro},
  {Rebecchi}, \& {Burgay}}]{B02}
{Burderi}, L., {Di Salvo}, T., {Stella}, L., {et~al.} 2002, \apj, 574, 930

\bibitem[{{Burderi} {et~al.}(2009){Burderi}, {Riggio}, {Di Salvo}, {Papitto},
  {Menna}, {D'A{\`i}}, \& {Iaria}}]{B09}
{Burderi}, L., {Riggio}, A., {Di Salvo}, T., {et~al.} 2009, \aap, 496, L17

\bibitem[{{Campana} {et~al.}(2008){Campana}, {Stella}, {Israel}, \&
  {D'Avanzo}}]{Cam08}
{Campana}, S., {Stella}, L., {Israel}, G., \& {D'Avanzo}, P. 2008, \apjl, 689,
  L129

\bibitem[{{Chakrabarty} {et~al.}(2003){Chakrabarty}, {Morgan}, {Muno},
  {Galloway}, {Wijnands}, {van der Klis}, \& {Markwardt}}]{C03}
{Chakrabarty}, D., {Morgan}, E.~H., {Muno}, M.~P., {et~al.} 2003, \nat, 424, 42

\bibitem[{{Chakrabarty} {et~al.}(2008){Chakrabarty}, {Swank}, {Markwardt}, \&
  {Smith}}]{C08}
{Chakrabarty}, D., {Swank}, J.~H., {Markwardt}, C.~B., \& {Smith}, E. 2008, The
  Astronomer's Telegram, 1660, 1

\bibitem[{{Deeter} {et~al.}(1981){Deeter}, {Boynton}, \&
  {Pravdo}}]{DeeBoyPrv81}
{Deeter}, J.~E., {Boynton}, P.~E., \& {Pravdo}, S.~H. 1981, \apj, 247, 1003

\bibitem[{{Di Salvo} \& {Burderi}(2003)}]{DSB03}
{Di Salvo}, T. \& {Burderi}, L. 2003, \aap, 397, 723

\bibitem[{{Di Salvo} {et~al.}(2008){Di Salvo}, {Burderi}, {Riggio},
  {et~al.}}]{DS08}
{Di Salvo}, T., {Burderi}, L., {Riggio}, A., {et~al.} 2008, \mnras, 389, 1851

\bibitem[{{Ek{\c s}i} {et~al.}(2005){Ek{\c s}i}, {Hernquist}, \&
  {Narayan}}]{Eks05}
{Ek{\c s}i}, K.~Y., {Hernquist}, L., \& {Narayan}, R. 2005, \apjl, 623, L41

\bibitem[{{Falanga} {et~al.}(2005){Falanga}, {Kuiper}, {Poutanen}, {Bonning},
  {Hermsen}, {Di Salvo}, {Goldoni}, {Goldwurm}, {Shaw}, \& {Stella}}]{Fal05}
{Falanga}, M., {Kuiper}, L., {Poutanen}, J., {et~al.} 2005, \aap, 444, 15

\bibitem[{{Galloway} {et~al.}(2005){Galloway}, {Markwardt}, {Morgan},
  {Chakrabarty}, \& {Strohmayer}}]{G05}
{Galloway}, D.~K., {Markwardt}, C.~B., {Morgan}, E.~H., {Chakrabarty}, D., \&
  {Strohmayer}, T.~E. 2005, \apjl, 622, L45

\bibitem[{{Ghosh} \& {Lamb}(1979)}]{GL79}
{Ghosh}, P. \& {Lamb}, F.~K. 1979, \apj, 234, 296

\bibitem[{{Hartman} {et~al.}(2011){Hartman}, {Galloway}, \&
  {Chakrabarty}}]{Hrt11}
{Hartman}, J.~M., {Galloway}, D.~K., \& {Chakrabarty}, D. 2011, \apj, 726, 26

\bibitem[{{Hartman} {et~al.}(2008){Hartman}, {Patruno}, {Chakrabarty},
  {Kaplan}, {Markwardt}, {Morgan}, {Ray}, {van der Klis}, \& {Wijnands}}]{H08}
{Hartman}, J.~M., {Patruno}, A., {Chakrabarty}, D., {et~al.} 2008, \apj, 675,
  1468

\bibitem[{{Hartman} {et~al.}(2009){Hartman}, {Patruno}, {Chakrabarty},
  {Markwardt}, {Morgan}, {van der Klis}, \& {Wijnands}}]{H09}
{Hartman}, J.~M., {Patruno}, A., {Chakrabarty}, D., {et~al.} 2009, \apj, 702,
  1673

\bibitem[{{Illarionov} \& {Sunyaev}(1975)}]{IllSun75}
{Illarionov}, A.~F. \& {Sunyaev}, R.~A. 1975, \aap, 39, 185

\bibitem[{{Jahoda} {et~al.}(2006){Jahoda}, {Markwardt}, {Radeva}, {Rots},
  {Stark}, {Swank}, {Strohmayer}, \& {Zhang}}]{Jah06}
{Jahoda}, K., {Markwardt}, C.~B., {Radeva}, Y., {et~al.} 2006, \apjs, 163, 401

\bibitem[{{Jonker} {et~al.}(2008){Jonker}, {Torres}, \& {Steeghs}}]{Jon08}
{Jonker}, P.~G., {Torres}, M.~A.~P., \& {Steeghs}, D. 2008, \apj, 680, 615

\bibitem[{{Leahy} {et~al.}(1983){Leahy}, {Darbro}, {Elsner}, {Weisskopf},
  {Kahn}, {Sutherland}, \& {Grindlay}}]{Leh83}
{Leahy}, D.~A., {Darbro}, W., {Elsner}, R.~F., {et~al.} 1983, \apj, 266, 160

\bibitem[{{Lyne} \& {Graham-Smith}(1990)}]{LynGsm90}
{Lyne}, A.~G. \& {Graham-Smith}, F. 1990, {Pulsar astronomy}, ed. {Cambridge,
  England and New York, Cambridge University Press (Cambridge Astrophysics
  Series, No.~16), 1990, 285 p.}

\bibitem[{{Markwardt} \& {Swank}(2008)}]{M08}
{Markwardt}, C.~B. \& {Swank}, J.~H. 2008, The Astronomer's Telegram, 1664, 1

\bibitem[{{Melatos} \& {Payne}(2005)}]{MelPay05}
{Melatos}, A. \& {Payne}, D.~J.~B. 2005, \apj, 623, 1044

\bibitem[{{Paizis} {et~al.}(2005){Paizis}, {Nowak}, {Wilms},
  {J-L.~Courvoisier}, {Ebisawa}, {Rodriguez}, \& {Ubertini}}]{Pzs05}
{Paizis}, A., {Nowak}, M.~A., {Wilms}, J., {et~al.} 2005, \aap, 444, 357

\bibitem[{{Papitto} {et~al.}(2005){Papitto}, {Menna}, {Burderi}, {Di Salvo},
  {D'Antona}, \& {Robba}}]{Pap05}
{Papitto}, A., {Menna}, M.~T., {Burderi}, L., {et~al.} 2005, \apjl, 621, L113

\bibitem[{{Patruno}(2010)}]{Pat10b}
{Patruno}, A. 2010, \apj, 722, 909

\bibitem[{{Patruno} {et~al.}(2010){Patruno}, {Altamirano}, \&
  {Messenger}}]{Pat10a}
{Patruno}, A., {Altamirano}, D., \& {Messenger}, C. 2010, \mnras, 403, 1426

\bibitem[{{Psaltis} \& {Chakrabarty}(1999)}]{PsaCha99}
{Psaltis}, D. \& {Chakrabarty}, D. 1999, \apj, 521, 332

\bibitem[{{Rupen} {et~al.}(2004){Rupen}, {Dhawan}, \& {Mioduszewski}}]{Rpn04}
{Rupen}, M.~P., {Dhawan}, V., \& {Mioduszewski}, A.~J. 2004, The Astronomer's
  Telegram, 364, 1

\bibitem[{{Shaw} {et~al.}(2005){Shaw}, {Mowlavi}, {Rodriguez}, {Ubertini},
  {Capitanio}, {Ebisawa}, {Eckert}, {Courvoisier}, {Produit}, {Walter}, \&
  {Falanga}}]{S05}
{Shaw}, S.~E., {Mowlavi}, N., {Rodriguez}, J., {et~al.} 2005, \aap, 432, L13

\bibitem[{{Spitkovsky}(2006)}]{Spt06}
{Spitkovsky}, A. 2006, \apjl, 648, L51

\bibitem[{{Thorne}(1980)}]{Tho80}
{Thorne}, K.~S. 1980, Reviews of Modern Physics, 52, 299

\bibitem[{{Torres} {et~al.}(2008){Torres}, {Jonker}, {Steeghs}, {Roelofs},
  {Bloom}, {Casares}, {Falco}, {Garcia}, {Marsh}, {Mendez}, {Miller},
  {Nelemans}, \& {Rodr{\'{\i}}guez-Gil}}]{T08}
{Torres}, M.~A.~P., {Jonker}, P.~G., {Steeghs}, D., {et~al.} 2008, \apj, 672,
  1079

\bibitem[{{Wagoner}(1984)}]{Wag84}
{Wagoner}, R.~V. 1984, \apj, 278, 345

\bibitem[{{Wijnands} \& {van der Klis}(1998)}]{WijvdK98}
{Wijnands}, R. \& {van der Klis}, M. 1998, \nat, 394, 344

\end{thebibliography}

\end{document}